\documentclass[twocolumn]{autart}    % Enable this line and disable the 
\usepackage{url}
\usepackage{amsmath,amssymb,amsfonts}
\usepackage{graphicx}          % Include this line if your 
\begin{document}

\begin{frontmatter}
%\runtitle{Insert a suggested running title}  % Running title for regular 
                                              % papers but only if the title  
                                              % is over 5 words. Running title 
                                              % is not shown in output.

\title{Monotonous Parameter Estimation of One Class of Nonlinearly Parameterized Regressions without Overparameterization\thanksref{footnoteinfo}} % Title, preferably not more 
                                                % than 10 words.

\thanks[footnoteinfo]{This research was financially supported in part by Grants Council of the President of the Russian Federation (project MD-1787.2022.4). The material in this paper was not presented at any IFAC 
meeting. Corresponding author A.~Glushchenko. Tel. +79102266946.}

\author[Moscow]{Anton Glushchenko}\ead{aiglush@upi.ru},    % Add the 
\author[Moscow]{Konstantin Lastochkin}\ead{lastconst@ipu.ru}               % e-mail address   % (ead) as shown

\address[Moscow]{Laboratory 7, V. A. Trapeznikov Institute of Control Sciences of Russian Academy of Sciences, Moscow, Russia}  % Please supply                                              
% here.

                             % keyword list or with the 
                                          % help of the Automatica 
                                          % keyword wizard

\begin{abstract}                          % Abstract of not more than 200 words.
The estimation law of unknown parameters vector ${\theta}$ is proposed for one class of nonlinearly parametrized regression equations $y\left( t \right) = \Omega \left( t \right)\Theta \left( \theta  \right)$. We restrict our attention to parametrizations that are widely obtained in practical scenarios when polynomials in $\theta$ are used to form $\Theta \left( \theta  \right)$. For them we introduce a new “linearizability” assumption that a mapping from overparametrized vector of parameters $\Theta \left( \theta  \right)$ to original one $\theta$ exists in terms of standard algebraic functions. Under such assumption and weak requirement of the regressor finite excitation, on the basis of dynamic regressor extension and mixing technique we propose a procedure to reduce the nonlinear regression equation to the linear parameterization without application of singularity causing operations and the need to identify the overparametrized parameters vector. As a result, an estimation law with exponential convergence rate is derived, which, unlike known solutions, \linebreak ({\it i}) does not require a strict {\emph{P}}-monotonicity condition to be met and \emph{a priori} information about $\theta$ to be known, ({\it ii}) ensures elementwise monotonicity for the parameter error vector. The effectiveness of our approach is illustrated with both academic example and 2-DOF robot manipulator control problem.
\end{abstract}

\end{frontmatter}

\section{Introduction}
In the majority of applications real technical systems have a limited number of significant physical parameters. At the same time, mathematical models of these systems, written in the state space or Euler-Lagrange form, are described by equations with overparameterization, {\emph{i.e.}} with a large number of new virtual parameters that are nonlinearly related to the original ones \cite{b3}, \cite{b2}.\linebreak

As far as classic methods of identification theory and adaptive control are concerned, each parameter of a mathematical model is considered to be unique and independent (decoupled) from the others. With increasing the system order and, as a result, the number of the above-mentioned virtual parameters, this leads to the well-known \cite{b3} shortcomings, which make it difficult to apply the basic estimation laws:
\begin{enumerate}
    \item [\textbf{S1.}] Slower convergence and stringent excitation conditions due to the need to solve the identification problem in a larger parameter space.
    \item [\textbf{S2.}] Necessity to apply projection operators for online estimation of the system physical parameters.
\end{enumerate}
To overcome these problems, it has been proposed \cite{b4}, \cite{b6}, \cite{b5},  \cite{b7} to take into account the relationship between the unknown parameters to design the estimation law. In \cite{b4} the dynamic regressor extension and mixing (DREM) technique is applied to “isolate the good mappings” from virtual to physical parameters and utilize the strong {\emph{P}}-monotonicity property \cite{b8} to achieve consistent parameter estimation for nonlinearly parameterized regressions. In case the regressor is non-square integrable, the solution \cite{b4} ensures asymptotic convergence of the parameter identification error. The requirement of strong {\emph{P}}-monotonicity has turned out to be strict enough for some applications, {\emph{e.g.}} composite control of Euler-Lagrange systems \cite{b5}, \cite{b2},  adaptive observation of windmill power coefficient \cite{b9}. For some, mainly polynomial mappings, it is possible to relax this condition using a special monotonizability assumption \cite{b9}, \cite{b6}, \cite{b5}. The relaxation mechanism is based on the search for a bijective substitution such that the new nonlinear mapping satisfies the strong {\emph{P}}-monotonicity condition. However, the solution from \cite{b6}, \cite{b5} has several key problems:
\begin{enumerate}
    \item [\textbf{P1.}]The convergence of the parametric error is guaranteed only for hardly validated non-square integrable regressors (Proposition 4 in \cite{b5});
    \item [\textbf{P2.}]Weak property of non-increasing norm of the parameter error vector is guaranteed, but not the elementwise monotonicity (Remark 8 in \cite{b5}).
    \item [\textbf{P3.}]The estimation law requires \emph{a priori} information about uncertainty parameters (for example, Lemma 2 in \cite{b5}).
    \item [\textbf{P4.}]The calculation of the system physical parameters from the obtained estimates can lead to singularities and sometimes requires application of projection operator (for example, see definition ${{\mathcal D}^I}$ in Lemma 2 of \cite{b5}). 
     \item [\textbf{P5.}] Due to \textbf{P2} and \textbf{P4} the transient behavior of parametric error is unpredictable, singularity may occur if we want to use ${{\mathcal D}^I}$.
\end{enumerate}
In a recent paper \cite{b7} a new estimation law has been proposed that solves \textbf{P1} and ensures exponential convergence of the parametric error when a more realistic for some practical scenarios condition of regressor finite excitation is satisfied. 

The motivation for this study is to solve all problems \textbf{P1-P5} for one class of nonlinearly parametrized regression equations (NLPRE).

\textbf{Notation and Definitions.} Further the following notation is used: $\left| . \right|$ is the absolute value, $\left\| . \right\|$ is the suitable norm of $(.)$, ${I_{n \times n}}=I_{n}$ is an identity $n \times n$ matrix, ${0_{n \times n}}$ is a zero $n \times n$ matrix, $0_{n}$ stands for a zero vector of length $n$, ${\rm{det}}\{.\}$ stands for a matrix determinant, ${\rm{adj}}\{.\}$ represents an adjoint matrix. Denote $\mathfrak{H}\left[ . \right]: = 1\;/\left( {p + k} \right)\left[ . \right]$ as a stable operator ($k > 0$ and $p: = d/dt$). For a mapping ${\mathcal F}{\rm{:\;}}{\mathbb{R}^n} \mapsto {\mathbb{R}^n}$  we denote its Jacobian by $\nabla_{x} {\mathcal F}\left( x \right) = \linebreak = {\textstyle{{\partial {\mathcal F}} \over {\partial x}}}\left( x \right)$. We also use the fact that for all (possibly singular) ${n \times n}$ matrices $M$ the following holds: ${\rm{adj}} \{M\} M  = {\rm{det}} \{M\}I_{n \times n}$.

\textbf{\emph{Definition.}}\emph{ A regressor $\omega \left( t \right)\in {\mathbb{R}^{n \times m}}$ is finitely exciting $(\omega \left( t \right) \in {\rm{FE}})$ over a time range $\left[ {t_r^ + {\rm{,\;}}{t_e}} \right]{\rm{}}$ if there exist $t_r^ +  \ge 0$, ${t_e} > t_r^ + $ and $\alpha $ such that the following inequality holds:
\begin{equation}\label{eq1}
\int\limits_{t_r^ + }^{{t_e}} {\omega \left( \tau  \right){\omega ^{\rm{T}}}\left( \tau  \right)d} \tau  \ge \alpha I_{n}{\rm{,}}
\end{equation}
where $\alpha \!>\! 0$ is the excitation level.}

\section{Problem Statement}

The following NLPRE is considered:
\begin{equation}\label{eq2}
y\left( t \right) = \Omega \left( t \right)\Theta \left( \theta  \right){\rm{,}}
\end{equation}
where $y\left( t \right) \in {\mathbb{R}^n}$,  $\Omega \left( t \right) \in {\mathbb{R}^{n \times p}}$ are measurable regressand and regressor, respectively, $\theta  \in {D_\theta } \subset {\mathbb{R}^q}$ is a vector of unknown time-invariant parameters, $\Theta {\rm{:\;}}{\mathbb{R}^q} \mapsto {\mathbb{R}^p}$ is a known mapping and $p > q$. The problem is to estimate parameters $\theta$ using $y(t)$ and $\Omega(t)$ such that:
\begin{subequations}  
\begin{equation}\label{eq3a}
\begin{array}{c}
\mathop {{\rm{lim}}}\limits_{t \to \infty } \left\| {\hat \theta \left( t \right) - \theta } \right\| = \mathop {{\rm{lim}}}\limits_{t \to \infty } \left\| {\tilde \theta \left( t \right)} \right\| = 0{\rm{ }}\left( {{\rm{exp}}} \right){\rm{,}}\\
\end{array}
\end{equation}
\begin{equation}\label{eq3b}
\begin{array}{c}
\forall {{\mathop{t}\nolimits} _a} \ge {t_b}{\rm{,\;}}\forall i \in \left\{ {1,{\rm{\;}}q} \right\}{\rm{\;}}\left| {{{\tilde \theta }_i}\left( {{t_a}} \right)} \right| \le \left| {{{\tilde \theta }_i}\left( {{t_b}} \right)} \right|{\rm{,}}
\end{array}
\end{equation}
\end{subequations}
where $\hat \theta \left( t \right)$ is an estimate of the unknown parameters, ${\tilde \theta _i}\left( t \right)$ is an estimation error of the \emph{i}$^{\rm th}$ parameter from $\theta$,  $\left(\rm{exp}\right)$ is an abbreviation for exponential rate of convergence. 

The feasibility conditions for the problem \eqref{eq3a} are
\begin{enumerate}
    \item [\textbf{FC1.}] $\Omega \left( t \right) \in {\rm{FE}}$, {\emph{i.e}.} condition of identifiability of an overparametrized parameters $\Theta \left( \theta  \right)$.
    \item [\textbf{FC2.}] ${D_\theta }{\rm{: = }}\Big\{ {\theta  \in {\mathbb{R}^{{q}}}{\rm{: de}}{{\rm{t}}}\Big\{ {{\nabla _\theta }{\psi}\left( \theta  \right)} \Big\} \ne  0} \Big\}$, where ${\psi}\left( \theta  \right) =\linebreak= {{\mathcal L}}\Theta \left( \theta  \right)\in {\mathcal{C}^{{k}}}{\rm{,}}$ {\emph{i.e.}} existence of inverse mapping  ${\mathcal F}{\rm{:\;}}{\mathbb{R}^q} \mapsto {\mathbb{R}^q}$ that reconstructs the unknown parameters $\theta  = {\mathcal F}\left( \psi  \right)$ from a "good" elements  $\psi \left( \theta \right)$ handpicked by $\mathcal{L}\in \mathbb{R}^{q \times p}$ from $\Theta \left( \theta  \right)$.
\end{enumerate}
When \textbf{FC1-FC2}\footnote{It should be understood that, if the inverse function ${\mathcal F}\left( \psi  \right)$ does not exist, then there is no way to obtain $\theta$ from $\Theta(\theta)$.}are met, then the parameters $\Theta \left( \theta  \right)$ can be obtained and recalculated into $\theta$ (possibly, only asymptotically). However the main contribution of this paper is to solve all problems \textbf{P1-P5}  and shortcomings \textbf{S1-S2} of existing solutions and consequently ensure elementwise monotonicity \eqref{eq3b} and obtain $\theta$ without estimation of $\Theta \left( \theta  \right)$ and substitution ${\mathcal F}\left( {\mathcal L}\hat{\Theta} \left( t  \right)  \right)$.

\section{Main Result}

To facilitate the proposed estimation design, in addition to \textbf{FC1-FC2} a class of mappings $\Theta \left( \theta  \right)$ and respective inverse functions ${\mathcal F}\left( \psi  \right)$, to which we restrict our attention, is defined in the following {\emph{linearizing}} assumption.

\textbf{\emph{Assumption 1.}}\emph{There exist ${\mathcal G}{\rm{:\;}}{\mathbb{R}^q} \!\mapsto\! {\mathbb{R}^{q \times q}}$, ${\mathcal S}{\rm{:\;}}{\mathbb{R}^q} \!\mapsto\! {\mathbb{R}^q}{\rm{,\;}}$ ${\Pi _\theta }{\rm{:\;}}{\mathbb{R}} \mapsto {\mathbb{R}^{q \times q}}{\rm{,}}$ ${{\mathcal T}_{\mathcal G}}{\rm{:\;}}{\mathbb{R}^{{\Delta _{\mathcal G}}}} \mapsto {\mathbb{R}^{q \times q}}{\rm{,}}$  ${{\mathcal T}_{\mathcal S}}{\rm{:\;}}{\mathbb{R}^{{\Delta _{\mathcal S}}}} \mapsto {\mathbb{R}^q}{\rm{,}}$ ${{\Xi _{\mathcal G}}}{\rm{:\;}}{\mathbb{R}} \mapsto {\mathbb{R}^{{{\Delta _{\mathcal G}}} \times q}}$, ${{\Xi _{\mathcal S}}}{\rm{:\;}}{\mathbb{R}} \mapsto {\mathbb{R}^{{{\Delta _{\mathcal S}}} \times q}}$ such that for all $\Delta \left( t \right) \in \mathbb{R}$  the following holds:} 
\begin{equation}\label{eq4}
\begin{array}{c}
{\mathcal S}\left( \psi  \right) = {\mathcal G}\left( \psi  \right){\mathcal F}\left( \psi  \right) = {\mathcal G}\left( \psi  \right)\theta {\rm{,}}\\
{\Pi _\theta }\left( \Delta  \right){\mathcal G}\left( \psi  \right) = {{\mathcal T}_{\mathcal G}}\left( {{\Xi _{\mathcal G}}\left( \Delta  \right)\psi } \right){\rm{,}}\\
{\Pi _\theta }\left( \Delta  \right) {{\mathcal S}\left( \psi  \right)}= {{\mathcal T}_{\mathcal S}}\left( {{\Xi _{\mathcal S}}\left( \Delta  \right)\psi } \right){\rm{,}}
\end{array}
\end{equation}
\emph{where}
\begin{gather*}
\begin{array}{c}
{\rm{det}}\left\{ {{\Pi _\theta }\left( \Delta  \right)} \right\} \ge {\Delta ^{{\ell _\theta }}}\left( t \right){\rm{,\;}}{\ell _\theta } \ge 1,{\rm{ rank}}\left\{ {{\mathcal G}\left( \psi  \right)} \right\} = q{\rm{,\;}}\\
{\Xi _{\left( . \right)}}\left( \Delta  \right) = {{\overline \Xi }_{\left( . \right)}}\left( \Delta  \right)\Delta \left( t \right) \in {\mathbb{R}^{{\Delta _{\left( . \right)}} \times q}}{\rm{, }}\\
{\Xi _{\left( . \right)}}_{ij}\left( \Delta  \right) = {c_{ij}}{\Delta ^{{\ell _{ij}}}}\left( t \right){\rm{,\;}}c_{ij} \in \left\{ {0,{\rm{ 1}}} \right\}{\rm{,\;}}{\ell _{ij}} \ge 1,
\end{array}
\end{gather*}
\emph{and all above mentioned mappings are known}\footnote{Assumption 1 is not restrictive and can be easily verified via direct inspection of mapping $\mathcal{F}(\psi)$.}.

Assumption 1 is met in case when  polynomials in $\theta$ are used to form $\Theta \left( \theta  \right)$ and consequently the inverse transform function ${\mathcal F}\left( \psi  \right)$ can be computed using algebraic functions. 

\textbf{Example.} For vector $\psi \left( \theta  \right) = col\left\{ {{\theta _1}{\theta _2}{\rm{ + }}\theta _1^2{\rm{,\;}}{\theta _2} + {\theta _1}} \right\}$  the mappings from \eqref{eq4} take the form:
\begin{equation}\label{eq5}
\begin{array}{c}
    {\mathcal S}\left( \psi  \right) = {\begin{bmatrix}
{{\psi _1}}\\
{\psi _2^2 - {\psi _1}}
\end{bmatrix}}{\rm{,\;}}{\mathcal G}\left( \psi  \right) =  {\begin{bmatrix}
{{\psi _2}}&0\\
0&{{\psi _2}}
\end{bmatrix} {\rm{,}}}\\
\begin{array}{c}
\Pi \left( \Delta  \right) = {\begin{bmatrix}
\Delta &0\\
0&{{\Delta ^2}}
\end{bmatrix}}{\rm{, }}\\
{\Xi _{\mathcal S}}\left( \Delta  \right) = {\begin{bmatrix}
\Delta &0\\
{{\Delta ^2}}&0\\
0&\Delta 
\end{bmatrix}}{\rm{,\;}}{\Xi _{\mathcal G}}\left( \Delta  \right) = {\begin{bmatrix}
0&\Delta \\
0&{{\Delta ^2}}
\end{bmatrix}}{\rm{, }}
\end{array}\\
{{\mathcal T}_{\mathcal G}}\left( {{\Xi _{\mathcal G}}\left( \Delta  \right)\psi } \right) = {\begin{bmatrix}
{\Delta {\psi _2}}&0\\
0&{{\Delta ^2}{\psi _2}}
\end{bmatrix}} {\rm{,  }}\\
\begin{array}{c}
{{\mathcal T}_{\mathcal S}}\left( {{\Xi _{\mathcal S}}\left( \Delta  \right)\psi } \right) = {\begin{bmatrix}
{{\psi _1}\Delta }\\
{{\Delta ^2}\psi _2^2 - {\Delta ^2}{\psi _1}}
\end{bmatrix}} {\rm{.}}\;\;\blacktriangledown\\
\end{array}
\end{array}\\
\end{equation}
Assumption 1 sets the conditions to obtain the following linearly parameterized regression equation from $\psi(\theta)$:
\begin{equation}\label{eq6}
{{\mathcal T}_{\mathcal S}}\left( {{\Xi _{\mathcal S}}\left( \Delta  \right)\psi } \right) = {{\mathcal T}_{\mathcal G}}\left( {{\Xi _{\mathcal G}}\left( \Delta  \right)\psi } \right)\theta {\rm{.}}
\end{equation}
Taking into consideration that the following equalities hold in accordance with Assumption 1:
\begin{equation}\label{eq7}
\begin{gathered}
    {\Xi _{\mathcal S}}\left( \Delta  \right) = {\overline \Xi _{\mathcal S}}\left( \Delta  \right)\Delta {\rm{,\;}}\\
    {\Xi _{\mathcal G}}\left( \Delta  \right) = {\overline \Xi _{\mathcal G}}\left( \Delta  \right)\Delta {\rm{,}}
\end{gathered}
\end{equation}
equation \eqref{eq6} is rewritten as:
\begin{equation}\label{eq8}
{{\mathcal T}_{\mathcal S}}\!\left( {{{\overline \Xi }_{\mathcal S}}\left( \Delta  \right){{\mathcal Y}_\psi }} \right) \!=\! {{\mathcal T}_{\mathcal G}}\left( {{{\overline \Xi }_{\mathcal G}}\left( \Delta  \right){{\mathcal Y}_\psi }} \right)\theta {\rm{,}}
\end{equation}
where $ {{\mathcal Y}_\psi }\left( t \right) = \Delta \left( t \right)\psi \left( \theta  \right)$ is the unmeasurable linear regression equation with respect to $ \psi \left( \theta  \right)$.

\textbf{Example (remainder).} For  $\psi \left( \theta  \right) = col\left\{ {{\theta _1}{\theta _2}{\rm{ + }}\theta _1^2{\rm{,\;}}{\theta _2} + {\theta _1}} \right\}$  the mappings from \eqref{eq8} take the form:
\begin{gather*}
\begin{array}{c}
{{\overline \Xi }_{\mathcal S}}\left( \Delta  \right) = {\begin{bmatrix}
1&0\\
\Delta &0\\
0&1
\end{bmatrix}}{\rm{,\;}}{{\overline \Xi }_{\mathcal G}}\left( \Delta  \right) = {\begin{bmatrix}
0&1\\
0&\Delta 
\end{bmatrix}}{\rm{, }}\\
{{\mathcal T}_{\mathcal G}}\left( {{\overline{\Xi} _{\mathcal G}}\left( \Delta  \right){{\mathcal Y}_\psi }} \right) = {\begin{bmatrix}
{{{{\mathcal Y}_{2\psi} }}}&0\\
0&{{\Delta}{{{\mathcal Y}_{2\psi}}}}
\end{bmatrix}} {\rm{,  }}\\
\begin{array}{c}
{{\mathcal T}_{\mathcal S}}\left( {{\overline{\Xi} _{\mathcal S}}\left( \Delta  \right){{\mathcal Y}_\psi }} \right) = {\begin{bmatrix}
{{{\mathcal Y}_{1\psi} }}\\
{{{\mathcal Y}^2_{2\psi}} - {\Delta}{{{{\mathcal Y}_{1\psi} }}}}
\end{bmatrix}} {\rm{.}}\;\;\blacktriangledown\\
\end{array}
\end{array}
\end{gather*}
Thus, if Assumption 1 is satisfied, having equation for ${{\mathcal Y}_\psi }\left( t \right)$ and the known mappings from \eqref{eq4} at hand, the regression equation with nonlinear parameterization \eqref{eq2} can be transformed into the new one with linear parameterization \eqref{eq8}. That is the reason why Assumption 1 is called “linearizing”. 

Using \eqref{eq2}, the regression equation with measurable ${{\mathcal Y}_\psi }\left( t \right){\rm{,\;}}\Delta \left( t \right) \ge 0$ could be obtained with the help of DREM procedure \cite{b4}. Towards this end, we introduce the following dynamic extension:
\begin{equation}\label{eq9}
\begin{array}{c}
\dot {\overline {y}}\left( t \right) = {e^{ - \sigma \left( {t - {t_0}} \right)}}{\Omega ^{\rm{T}}}\left( t \right)y\left( t \right){\rm{,\;}}\overline y\left( {{t_0}} \right) = {0_p},\\
\dot {\overline {\Omega}} \left( t \right) = {e^{ - \sigma \left( {t - {t_0}} \right)}}{\Omega ^{\rm{T}}}\left( t \right)\Omega \left( t \right){\rm{,\;}}\overline \Omega \left( {{t_0}} \right) = {0_{p \times p}},
\end{array}
\end{equation}
and apply a mixing procedure to $\overline{y}(t)$:
\begin{equation}\label{eq10}
\begin{array}{c}
{{\mathcal Y}_\psi }\left( t \right) = \Delta \left( t \right)\psi \left( \theta  \right){\rm{, }}\\
{{\mathcal Y}_\psi }\left( t \right)\!{\rm{:}} =\! {\mathcal L}{\rm{adj}}\left\{ {\overline \Omega \left( t \right)} \right\}\overline y\left( t \right){\rm{,\;}}\Delta \left( t \right)\!{\rm{:}}\! =\! {\rm{det}}\left\{ {\overline \Omega \left( t \right)} \right\}.
\end{array}
\end{equation}
The following proposition has been proved in \cite{b15}, \cite{b14} for the scalar regressor $\Delta \left( t \right)$ obtained by \eqref{eq9} and \eqref{eq10}.

\emph{ \textbf{Proposition 1.} If $\Omega \left( t \right) \in {\rm{FE}}$, then for all $t \ge {t_e} \; \Delta \left( t \right) \ge \linebreak \ge {\Delta _{LB}} > 0.$}

So, the signals ${{\mathcal T}_{\mathcal S}}\left( {{{\overline \Xi }_{\mathcal S}}\left( \Delta  \right){{\mathcal Y}_\psi }} \right){\rm{,\;}}{{\mathcal T}_{\mathcal G}}\left( {{{\overline \Xi }_{\mathcal G}}\left( \Delta  \right){{\mathcal Y}_\psi }} \right)$ can be computed through equations \eqref{eq9} and \eqref{eq10}.  Then the mixing procedure is applied \emph{a novo}:
\begin{equation}\label{eq11}
\begin{array}{c}
{{\mathcal Y}_\theta }\left( t \right) = {\mathcal M}\left( t \right)\theta {\rm{, }}\\
{{\mathcal Y}_\theta }\left( t \right)\!{\rm{:}}\! =\! {\rm{adj}}\!\left\{ {{{\mathcal T}_{\mathcal G}}\left( {{{\overline \Xi }_{\mathcal G}}\left( \Delta  \right){{\mathcal Y}_\psi }} \right)} \right\}{{\mathcal T}_{\mathcal S}}\left( {{{\overline \Xi }_{\mathcal S}}\left( \Delta  \right){{\mathcal Y}_\psi }} \right) {\rm{,}}\\
{\rm{ }}{\mathcal M}\left( t \right){\rm{:}} = {\rm{det}}\left\{ {{{\mathcal T}_{\mathcal G}}\left( {{{\overline \Xi }_{\mathcal G}}\left( \Delta  \right){{\mathcal Y}_\psi }} \right)} \right\}.
\end{array}
\end{equation}
Having the linear regression equation \eqref{eq11} at hand, the estimation law to identify the unknown parameters is introduced based on standard gradient descent method:
\begin{equation}\label{eq12}
\dot{ \hat {\theta}} \left( t \right) = \dot{ \tilde {\theta}} \left( t \right) =  - \gamma {\mathcal M}\left( t \right)\left( {{\mathcal M}\left( t \right)\hat \theta \left( t \right) - {{\mathcal Y}_\theta }\left( t \right)} \right),
\end{equation}
where $\gamma > 0$ is an adaptive gain, $\hat {\theta}(t_{0})=\hat{\theta}_{0}$ is an initial condition.

The properties of the law \eqref{eq12} are considered in the following theorem.

\emph{\textbf{Theorem 1.} If} \textbf{FC1}-\textbf{FC2} {\emph{and Assumption 1 are met, then goals \eqref{eq3a} and \eqref{eq3b} are achieved.}}

\emph{Proof.} The solution of the differential equation \eqref{eq12} for all $t \ge {t_0}$  is written as:
\begin{equation}\label{eq13}
{\tilde \theta _i}\left( t \right) = {e^{ - \gamma \int\limits_{{t_0}}^t {{{\mathcal M}^2}\left( \tau  \right)d\tau } }}{\tilde \theta _i}\left( {{t_0}} \right){\rm{,}}
\end{equation}
from which $ \forall {{\mathop{t}\nolimits} _a} \ge {t_b}{\rm{,\;}}\forall i \in \left\{ {1,{\rm{ }}q} \right\}{\rm{ }}\left| {{{\tilde \theta }_i}\left( {{t_a}} \right)} \right| \le \left| {{{\tilde \theta }_i}\left( {{t_b}} \right)} \right|$.

Following Assumption 1, it holds that $ {\rm{det}}\left\{ {{\Pi _\theta }\left( \Delta  \right)} \right\} \ge \linebreak \ge {\Delta ^{{\ell _\theta }}}\left( t \right){\rm{, rank}}\left\{ {{\mathcal G}\left( \psi  \right)} \right\} = q$, and $ \forall t \ge {t_e}{\rm{\;}} \Delta \left( t \right) \ge {\Delta _{LB}} > 0$ also holds owing to Proposition 1, then for all $t \ge {t_e}$ we can write the following expression for ${\mathcal M}\left( t \right)$:
\begin{equation}\label{eq14}
\begin{array}{c}
\left| {{\mathcal M}} \right| = \left| {{\rm{det}}\left\{ {{{\mathcal T}_{\mathcal G}}\left( {{\Xi _{\mathcal G}}\left( \Delta  \right){{\mathcal Y}_\psi }} \right)} \right\}} \right| = \left| {{\rm{det}}\left\{ {{\Pi _\theta }\left( \Delta  \right)} \right\}} \right| \cdot \\
 \cdot \left| {{\rm{det}}\left\{ {{\mathcal G}\left( \psi  \right)} \right\}} \right| \ge \Delta _{{{LB}}}^{{\ell _\theta }}\left| {{\rm{det}}\left\{ {{\mathcal G}\left( \psi  \right)} \right\}} \right| > 0,
\end{array}
\end{equation}
which, in its turn, allows one to rewrite the solution \eqref{eq13} for all $t \ge {t_e}$ as:
\begin{equation}\label{eq15}
\left| {{{\tilde \theta }_i}\left( t \right)} \right| \le {e^{ - \gamma \Delta _{{{LB}}}^{2{\ell _\theta }}{\rm{de}}{{\rm{t}}^2}\left\{ {{\mathcal G}\left( \psi  \right)} \right\}\left( {t - {t_e}} \right)}}\left| {{{\tilde \theta }_i}\left( {{t_0}} \right)} \right|{\rm{,}}
\end{equation}
from which it follows that $\mathop {{\rm{lim}}}\limits_{t \to \infty } \left\| {\tilde \theta \left( t \right)} \right\| = 0{\rm{ }}\left( {{\rm{exp}}} \right)$.

\emph{This completes the proof of Theorem 1.}

Therefore, if the mapping ${\mathcal F}\left( \psi  \right)$ satisfy the premises of Assumption 1, then, in accordance with the extension \eqref{eq9} and mixing procedures \eqref{eq10} and \eqref{eq11}, the estimation law \eqref{eq12} can be designed ensuring that the goals \eqref{eq3a} and \eqref{eq3b} are achieved. Note that, in contrast to \cite{b7}, in addition to properties \eqref{eq3a} and \eqref{eq3b} the proposed law does not use \emph{a priori} information about low and upper bounds of parameters (\textbf{P3}) in design procedure\footnote{It should be noted that the proposed law requires only knowledge that $\theta$ lies in the safe domain $D_{\theta}$ from \textbf{FC2}.} and does not include singularity causing division operations (\textbf{P4-P5}).

\section{Numerical Experiment}
\subsection{Academic example}
Using an academic example, the proposed identification method has been compared with the gradient law and the one proposed in \cite{b5}. The regressor and the mapping were defined as follows:
\begin{equation}\label{eq16}
\Omega \left( t \right) = {{\begin{bmatrix}
  {{e^{ - t}}} \\ 
  {\sin \left( t \right)} \\ 
  1 
\end{bmatrix}}^{\text{T}}}{\text{, }}\Theta \left( \theta  \right) =  {\begin{bmatrix}
  {{\theta _1}{\theta _2}{\text{ + }}\theta _1^2} \\ 
  {{\theta _2} + {\theta _1}} \\ 
  {{\text{cos}}\left( {{\theta _1}} \right)} 
\end{bmatrix}}{\text{, }}    
\end{equation}
where the \textbf{FC1} was met, and the premises of \textbf{FC2} were satisfied in case $\theta \in {D_\theta }{\rm{: = }}\Big\{ {\theta  \in {\mathbb{R}^{{q}}}{\rm{:\;}}{\theta _2} \ne  - {\theta _1}} \Big\}$.

According to the proposed approach, the matrix $\mathcal{L}= \linebreak =\begin{bmatrix} I_{2}&0_{2} \end{bmatrix}$ to implement the mixing procedure \eqref{eq10} was introduced, and the mappings from Assumption 1 that were necessary to implement \eqref{eq12} were defined as in the above-given example (see \eqref{eq5}).

In accordance with the “monotonizability Assumption”, the following change of variables was introduced to implement the estimation law from \cite{b5}:
\begin{equation}\label{eq17}
\begin{gathered}
  \eta  = \mathcal{D}\left( \theta  \right) = col\left\{ {{\theta _1}{\text{, }}{\theta _1} + {\theta _2}} \right\}{\text{,}} \\ 
  \theta  = {\mathcal{D}^I}\left( \eta  \right) = col\left\{ {{\eta _1}{\text{, }}{\eta _2} - {\eta _1}} \right\}{\text{,}} \\ 
\end{gathered}
\end{equation}
which allowed one to rewrite $\Theta(\theta)$ as $\left( {\Theta  \circ {\mathcal{D}^I}} \right)\left( \eta  \right) = \linebreak = col\left\{ {{\eta _1}{\eta _2}{\text{, }}{\eta _2}{\text{, cos}}\left( {{\eta _1}} \right)} \right\}$ and ensure that there existed $\rho  > 0$ such that the strong {\emph{P}}-monotonicity condition \cite{b8} for mapping $\mathcal{W} \left( \eta  \right) = \mathcal{L}\left( {\Theta  \circ {\mathcal{D}^I}} \right)\left( \eta  \right)=\left( {\psi  \circ {\mathcal{D}^I}} \right)\left( \eta  \right)$:
\begin{equation}\label{eq18}
\begin{gathered}
  {\left( {a - b} \right)^{\text{T}}}P\left( {\mathcal{W} \left( a \right) - \mathcal{W} \left( b \right)} \right) \geqslant \rho {\left| {a - b} \right|^2} > 0{\text{,}} \\ 
  \forall a{\text{, }}b \in {\mathbb{R}^2}{\text{, }}a \ne b{\text{, }} \\ 
\end{gathered}
\end{equation}
was met for $P = {\begin{bmatrix}
  \kappa &0 \\ 
  0&1 
\end{bmatrix}}{\text{, }}\kappa  \geqslant \tfrac{{\theta _{1{\text{max}}}^2}}{{4\left( {{\theta _{1{\text{min}}}} + {\theta _{2{\text{min}}}}} \right)}}$.

According to \cite{b5} and using \eqref{eq10}, the parameter estimation law was rewritten as:
\begin{equation}\label{eq19}
\begin{gathered}
  \hat \theta \left( t \right) = {\mathcal{D}^I}\left( {\hat \eta } \right){\text{,}} \\
  \dot {\hat \eta} \left( t \right) = {\gamma _\eta }P\Delta \left( t \right)\left( {{\mathcal{Y}_\psi }\left( t \right) - \Delta \left( t \right)\mathcal{W} \left( {\hat \eta } \right)} \right). \\ 
\end{gathered}
\end{equation}
The classic gradient-based estimation law was defined as:
\begin{equation}\label{eq20}
\begin{gathered}
  \hat \theta \left( t \right) = {\begin{bmatrix}
  {\tfrac{{{{\hat \Theta }_1}\left( t \right)}}{{{{\hat \Theta }_2}\left( t \right)}}} \\ 
  {{{\hat \Theta }_2}\left( t \right) - \tfrac{{{{\hat \Theta }_1}\left( t \right)}}{{{{\hat \Theta }_2}\left( t \right)}}} 
\end{bmatrix}}{\text{,}} \\
  \dot {\hat \Theta} \left( t \right) =  - \Gamma {\Omega ^{\text{T}}}\left( t \right)\left( {\Omega \left( t \right)\hat \Theta \left( t \right) - y\left( t \right)} \right). \\ 
\end{gathered}
\end{equation}
It should be noted that in contrast to \eqref{eq12}, the law \eqref{eq19} required information about the low bounds ${\theta _{1{\text{min}}}},\;{\theta _{2{\text{min}}}}$, while the law \eqref{eq20} included the division operation. To conduct the experiment, the unknown parameters $\theta$, parameters of filters \eqref{eq9} and laws \eqref{eq12}, \eqref{eq19}, \eqref{eq20} were set as follows:
\begin{equation}\label{eq21}
\begin{gathered}
  {\theta _1} = 1,{\text{ }}{\theta _2} = 2, \\ 
  \sigma  = 1,{\text{ }}\gamma  = {10^{13}}{\text{, }}{\gamma _\eta } = {10^5}{\text{, }}\Gamma  = 10{I_2}{\text{, }}\kappa  = 10, \\ 
  \hat \theta \left( 0 \right) = \hat \eta \left( 0 \right) = {0_2}{\text{, }}\hat \Theta \left( 0 \right) = {\begin{bmatrix}
  0&1&0 
\end{bmatrix}}. \\ 
\end{gathered}
\end{equation}
The initial conditions for \eqref{eq20} were chosen by trial and error so that to meet the condition ${\hat \Theta _2}\left( t \right) \ne 0$. Figure 1 depicts the transients of the estimates obtained with the help of \eqref{eq12}, \eqref{eq19}, and \eqref{eq20}.

   \begin{figure}[h]
      \centering
      \includegraphics[scale=0.67]{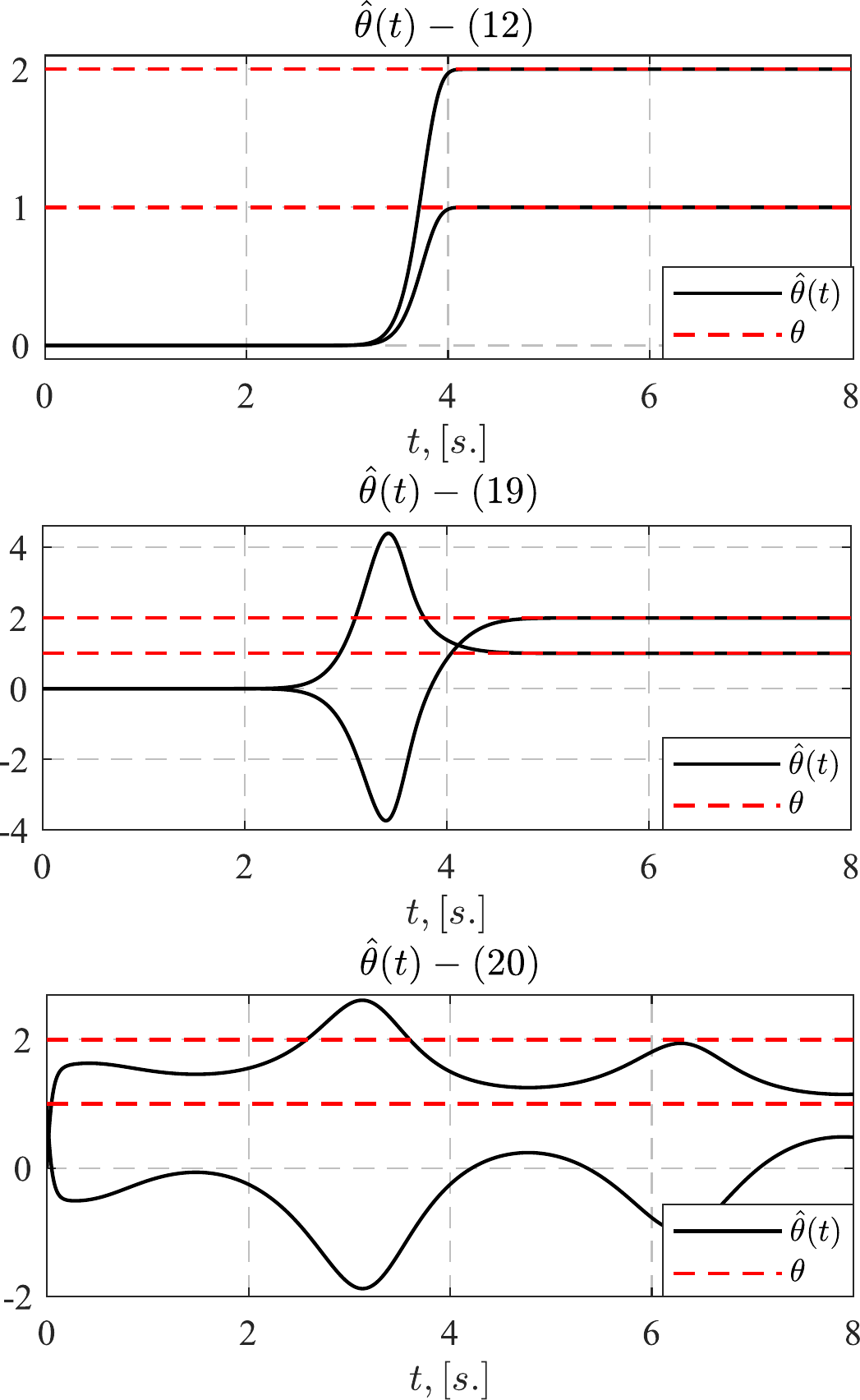}
      \caption{Transient behavior of $\hat \theta \left( t \right)$}
      \label{Figure1} 
      \end{figure}
      
Estimates obtained with the laws \eqref{eq12} and \eqref{eq19} exponentially converged to the true values, since the condition $\Omega \left( t \right) \in {\text{FE}}$ was met. At the same time, the estimates by \eqref{eq20} did not converge to true values since $\Omega \left( t \right) \notin {\text{PE}}$. The simulation result confirmed that the goal \eqref{eq3a} and \eqref{eq3b} was achieved and demonstrated the advantages of the proposed solution in comparison with both the classic gradient identifier with overparameterization \eqref{eq20} and the law \eqref{eq19} from \cite{b5}.

\subsection{2-DOF robot manipulator}
A problem of adaptive control of a 2-DOF robot manipulator with uncertainty has been considered:
\begin{equation}\label{eq22}
\begin{gathered}
  M\left( q \right)\ddot q + C\left( {q{\text{, }}\dot q} \right)\dot q + \nabla U\left( q \right) = u{\text{,}}  
\end{gathered}
\end{equation}
\begin{gather*}
     M\!\left( q \right)\! =\! {\begin{bmatrix}
  {{\Theta _1}\!\left( \theta  \right)\! +\! 2{\Theta _2}\!\left( \theta  \right)\cos\! \left( {{q_2}} \right)}&{{\Theta _3}\!\left( \theta  \right)\! +\! {\Theta _2}\!\left( \theta  \right)\cos \left( {{q_2}} \right)} \\ 
  {{\Theta _3}\!\left( \theta  \right)\! +\! {\Theta _2}\!\left( \theta  \right)\cos \!\left( {{q_2}} \right)}&{{\Theta _3}\!\left( \theta  \right)} 
\end{bmatrix}}{\text{,}} \\ 
  C\left( {q{\text{, }}\dot q} \right)\! =\! {\begin{bmatrix}
  { - {\Theta _2}\left( \theta  \right)\sin \left( {{q_2}} \right){{\dot q}_2}}&{ - {\Theta _2}\left( \theta  \right)\sin \left( {{q_2}} \right)\left( {{{\dot q}_1}\! +\! {{\dot q}_2}} \right)} \\ 
  {{\Theta _2}\left( \theta  \right)\sin \left( {{q_2}} \right){{\dot q}_1}}&0 
\end{bmatrix}}{\text{,}} \\ 
  \nabla U\left( q \right) = {\begin{bmatrix}
  {{\Theta _4}\left( \theta  \right)g\cos \left( {{q_1} + {q_2}} \right) + {\Theta _5}\left( \theta  \right)g\cos \left( {{q_1}} \right)} \\ 
  {{\Theta _4}\left( \theta  \right)g\cos \left( {{q_1} + {q_2}} \right)} 
\end{bmatrix}}{\text{,}} \\ 
  \Theta \left( \theta  \right) = {\begin{bmatrix}
  {\theta _2^2{\theta _4} + \theta _1^2\left( {{\theta _3} + {\theta _4}} \right)} \\ 
  {{\theta _1}{\theta _2}{\theta _4}} \\ 
  {\theta _2^2{\theta _4}} \\ 
  {{\theta _2}{\theta _4}} \\ 
  {{\theta _1}\left( {{\theta _3} + {\theta _4}} \right)} 
\end{bmatrix}}{\text{,}}
\end{gather*}
where $q \in {\mathbb{R}^2}$ is the vector of generalized coordinates, $u \in {\mathbb{R}^2}$ is the control vector, $M{\text{: }}{\mathbb{R}^2} \mapsto {\mathbb{R}^{2 \times 2}}$ is the generalized inertia matrix, which is positive definite and assumed to be bounded, $C{\text{: }}{\mathbb{R}^2} \times {\mathbb{R}^2} \mapsto {\mathbb{R}^{2 \times 2}}$ represents the Coriolis and centrifugal forces matrix, $U{\text{: }}{\mathbb{R}^2} \mapsto \mathbb{R}$ is the potential energy function.

The goal was stated as $\mathop {{\text{lim}}}\limits_{t \to \infty } {\text{ }}col\left\{ {\tilde q{\text{, }}\dot {\tilde q}} \right\} = 0,$ where \linebreak $\tilde q = q - {q_ * }$ is state tracking error, and ${q_ * }$ is a reference trajectory. Certainty equivalence Slotine-Li controller \cite{b16} that ensured achievement of the above-mentioned goal had the form \cite{b5}:
\begin{equation}\label{eq23}
\begin{gathered}
  u = W\left( {q{\text{, }}\dot q{\text{, }}t} \right)\Theta \left( {\hat \theta } \right) + {K_1}s{\text{, }}s = \dot {\tilde q} + {K_2}\tilde q{\text{,}} \\ 
  W\left( {q{\text{, }}\dot q{\text{, }}t} \right) = {\begin{bmatrix}
  {{W_{11}}}&{{W_{12}}}&{{W_{13}}}&{{W_{14}}}&{{W_{15}}} \\ 
  {{W_{21}}}&{{W_{22}}}&{{W_{23}}}&{{W_{24}}}&{{W_{25}}} 
\end{bmatrix}}{\text{,}} \\ 
\end{gathered}
\end{equation}
with ${W_{11}} = {\ddot q_{r1}}{\text{, }}{W_{12}} = \cos \left( {{q_2}} \right)\left( {2{{\ddot q}_{r1}} + {{\ddot q}_{r2}}} \right) - sin\left( {{q_2}} \right)\cdot \linebreak \cdot \left( {{{\dot q}_2}{{\dot q}_{r1}}\! +\! \left( {{{\dot q}_1}\! +\! {{\dot q}_2}} \right){{\dot q}_{r2}}} \right){\text{, }}$ ${W_{14}}\! =\! {W_{24}} \!=\! g\cos \left( {{q_1}\! +\! {q_2}} \right){\text{, }}$  ${W_{13}} = {\ddot q_{r2}}{\text{, }}{W_{15}} = g\cos \left( {{q_1}} \right){\text{, }}{W_{21}} = {W_{25}} = 0,{\text{ }}\linebreak {W_{22}} = \cos \left( {{q_2}} \right){\ddot q_{r1}} + \sin \left( {{q_2}} \right){\dot q_1}{\dot q_{r1}},$
${W_{23}} = {\ddot q_{r1}} + {\ddot q_{r2}}$ where ${\dot q_r} = {\dot q_ * } - {K_2}\tilde q$.

The estimates of the unknown parameters $\theta $ with exponential or asymptotic rate of convergence were required to implement \eqref{eq23}. Using measurable signals $q{\text{, }}\dot q$ and $\tau $ and the results of Proposition 7 from \cite{b5}, the regression model \eqref{eq2} was parametrized as follows:
\begin{equation}\label{eq24}
y\left( t \right)\! =\! \mathfrak{H}\left[ u \right]{\text{, }}\Omega \left( t \right) \!=\! \mathfrak{H} {\begin{bmatrix}
  {{\Omega _{11}}}&{{\Omega _{12}}}&{{\Omega _{13}}}&{{\Omega _{14}}}&{{\Omega _{15}}} \\ 
  {{\Omega _{21}}}&{{\Omega _{22}}}&{{\Omega _{23}}}&{{\Omega _{24}}}&{{\Omega _{25}}} 
\end{bmatrix}}{\text{,}}
\end{equation}
and ${\Omega _{11}}\! =\! p{\dot q_1}{\text{, }}{\Omega _{12}}\! =\! p\cos \left( {{q_2}} \right)\left( {2{{\dot q}_1} + {{\dot q}_2}} \right){\text{, }}{\Omega _{13}}\! =\! p{\dot q_2}{\text{, }}\linebreak {\Omega _{14}} = {\Omega _{24}} = {W_{14}}{\text{, }}{\Omega _{15}} = {W_{15}}{\text{, }}$
${\Omega _{21}} = {\Omega _{25}} = 0,{\text{ }} \linebreak {\Omega _{22}} = p\cos \left( {{q_2}} \right){\dot q_1} + \sin \left( {{q_2}} \right)\left( {\dot q_1^2 + {{\dot q}_1}{{\dot q}_2}} \right){\text{, }}{\Omega _{23}} = p\left( {{{\dot q}_1} + {{\dot q}_2}} \right)$.
In accordance with the “monotonizability Assumption”, the following change of variables was introduced to implement the identification law from \cite{b5}:
\begin{equation}\label{eq25}
\begin{gathered}
  \eta  = \mathcal{D}\left( \theta  \right) = col\left\{ {{\theta _1}{\text{, }}{\theta _2}{\text{, }}{\theta _2}{\theta _4}{\text{, }}{\theta _1}\left( {{\theta _3} + {\theta _4}} \right)} \right\}{\text{,}} \\ 
  \theta  = {\mathcal{D}^I}\left( \eta  \right) = col\left\{ {{\eta _1}{\text{, }}{\eta _2}{\text{, }}\tfrac{{{\eta _4}}}{{{\eta _1}}} - \tfrac{{{\eta _3}}}{{{\eta _2}}}{\text{, }}\tfrac{{{\eta _3}}}{{{\eta _2}}}} \right\}{\text{,}} \\ 
\end{gathered}
\end{equation}
which allowed one to rewrite $\Theta \left( \theta  \right)$ as $\left( {\Theta  \circ {\mathcal{D}}^I} \right)\left( \eta  \right) = \linebreak = col\left\{ {{\eta _2}{\eta _3} + {\eta _1}{\eta _4}{\text{, }}{\eta _1}{\eta _3}{\text{, }}{\eta _2}{\eta _3}{\text{, }}{\eta _3}{\text{, }}{\eta _4}} \right\}$ and ensure that there existed a constant $\rho  > 0$ such that the strong {\emph{P}}-monotonicity condition \cite{b8} for mapping $\mathcal{W}\left( \eta  \right) = C\left( {\Theta  \circ {\mathcal{D}}^I} \right)\left( \eta  \right)$:
\begin{equation}\label{eq26}
\begin{gathered}
  {\left( {a - b} \right)^{\text{T}}}P\left( {\mathcal{W}\left( a \right) - \mathcal{W}\left( b \right)} \right) \geqslant \rho {\left| {a - b} \right|^2} > {\text{0}}{\text{,}} \\ 
  \forall a{\text{, }}b \in {\mathbb{R}^4}{\text{, }}a \ne b{\text{, }} \\ 
\end{gathered}
\end{equation}
was met for
\begin{displaymath}
\begin{gathered}
C = {\begin{bmatrix}
  {{I_4}}&{{0_4}} 
\end{bmatrix}}{\begin{bmatrix}
  {{0_4}}&{{I_4}} \\ 
  1&{{0_{1 \times 4}}} 
\end{bmatrix}}{\text{, }}P = diag\left\{ {\kappa {\text{, }}\kappa {\text{, 0}}{\text{, 0}}} \right\}{\text{, }}\\
\kappa  \geqslant \tfrac{1}{{4\theta _4^m}}\left[ {\theta _2^M + \tfrac{{{{\left( {\theta _1^M} \right)}^2}}}{{\theta _2^m}}} \right].
\end{gathered}
\end{displaymath}
According to \cite{b5} and using \eqref{eq9}, the estimation law was defined as:
\begin{equation}\label{eq27}
\begin{gathered}
  \hat \theta \left( t \right) = {\mathcal{D}^I}\left( {\hat \eta } \right){\text{,}} \\
  \dot {\hat \eta} \!\left( t \right)\! =\! {\gamma _\eta }P\Delta\! \left( t \right)\!\left( {C{\text{adj}}\left\{ {\overline \Omega \left( t \right)} \right\}\overline y\left( t \right) \!-\! \Delta \left( t \right)\mathcal{W}\left( {\hat \eta } \right)} \right). \\ 
\end{gathered}
\end{equation}
Following the proposed method of identification, the vector $\psi\left( \theta  \right)$ from \textbf{FC2} and the mappings from Assumption 1 took the form:
\begin{equation}\label{eq28}
\begin{gathered}
  \psi\left( \theta  \right)  =  {\begin{bmatrix}
  {{\Theta _1}\left( \theta  \right)}&{{\Theta _2}\left( \theta  \right)}&{{\Theta _3}\left( \theta  \right)}&{{\Theta _5}\left( \theta  \right)} 
\end{bmatrix}} {\text{,}} \\
\mathcal{G}\!\left( \psi  \right)\! =\! diag\!\left\{ {{\psi _4}{\text{, }}{\psi _4}{\psi _2}{\text{, }}{{\left( {{\psi _1} \!-\! {\psi _3}} \right)}^2}\!{\psi _3}{\text{, }}{{\left( {{\psi _1} \!-\! {\psi _3}} \right)}^2}{\psi _3}} \!\right\}\!{\text{,}}\\
\mathcal{S}\left( \psi  \right) = col\left\{ {{\psi _1} - {\psi _3}{\text{, }}\left( {{\psi _1} - {\psi _3}} \right){\psi _3}{\text{, }}}\right.
  \\
  \left.{\left( {{\psi _1}\! -\! {\psi _3}} \right){\psi _3}\psi _4^2 - \psi _4^2\psi _2^2{\text{, }}\psi _4^2\psi _2^2} \right\}{\text{,}}
  \\
  {\mathcal{T}_\mathcal{G}}\left( {{{\overline \Xi }_\mathcal{G}}\left( \Delta  \right){\mathcal{Y}_\psi }} \right) = diag\left\{ {\mathcal{Y}_{4\psi }}{\text{, }}{\mathcal{Y}_{4\psi }}{\mathcal{Y}_{2\psi }}{\text{, }}\right.
  \\
  \left.{{\left( {{\mathcal{Y}_{1\psi }} - {\mathcal{Y}_{3\psi }}} \right)}^2}\Delta {\mathcal{Y}_{3\psi }}{\text{, }}{{\left( {{\mathcal{Y}_{1\psi }} - {\mathcal{Y}_{3\psi }}} \right)}^2}\Delta {\mathcal{Y}_{3\psi }} \right\}{\text{,}}
  \\
  {\mathcal{T}_\mathcal{S}}\left( {{{\overline \Xi }_\mathcal{S}}\left( \Delta  \right){\mathcal{Y}_\psi }} \right)\! =\! col\left\{ {\mathcal{Y}_{1\psi }} \!-\! {\mathcal{Y}_{3\psi }}{\text{, }}\left( {{\mathcal{Y}_{1\psi }} \!-\! {\mathcal{Y}_{3\psi }}} \right){\mathcal{Y}_{3\psi }}{\text{, }}\right.
  \\
  \left.\left( {{\mathcal{Y}_{1\psi }} - {\mathcal{Y}_{3\psi }}} \right){\mathcal{Y}_{3\psi }}\mathcal{Y}_{4\psi }^2 - \mathcal{Y}_{4\psi }^2\mathcal{Y}_{2\psi }^2{\text{, }}\mathcal{Y}_{4\psi }^2\mathcal{Y}_{2\psi }^2 \right\}{\text{.}}
\end{gathered}
\end{equation}
Note that, unlike \eqref{eq12}, the law \eqref{eq27} requires information about the bounds ${\theta _1} \leqslant \theta _1^M{\text{, }}\theta _2^m \leqslant {\theta _2} \leqslant \theta _2^M{\text{, }}\theta _4^m \leqslant {\theta _4}$ and uses the singularity burden division operation in the mapping ${\mathcal{D}^I}\left( {\hat \eta } \right)$. To conduct the experiment, the unknown parameters $\theta$, parameters of the control law \eqref{eq23}, filters \eqref{eq9} and laws \eqref{eq12}, \eqref{eq27} were set as follows:
\begin{equation}\label{eq32}
\begin{gathered}
  {\theta _1} = 0.7,{\text{ }}{\theta _2} = 0.8,{\text{ }}{\theta _3} = 1.5,{\text{ }}{\theta _4} = 0.5,{\text{ }}g = 9.{\text{8}}{\text{,}} \\ 
  {K_1} = 3{I_2}{\text{, }}{K_2} = {I_2}{\text{, }}\sigma  = 1,{\text{ }}\kappa  = 10, \\ 
  {{\hat \eta }_i}\left( 0 \right) = 0.1,{\text{ }}\hat \theta \left( 0 \right) = {\mathcal{D}^I}\left( {\hat \eta \left( 0 \right)} \right){\text{, }}
  \\
\gamma {\rm{:}} = {\textstyle{{10} \over {1 + {{\mathcal M}^2}\left( t \right)}}}{\rm{,\;}}{\gamma _\eta }{\rm{:}} = {\textstyle{5 \over {1 + {\Delta ^2}\left( t \right)}}}.
\end{gathered}
\end{equation}
It is worth mentioning that the applicability and safety of use of time-varying adaptive gain in certainty equivalence indirect control problem was shown in, for instance, Proposition 6 from \cite{b5}. So the above-presented proof of Theorem is correct \emph{mutatis mutandis} for this simulation example.

Figure 2 presents the transients of both estimates $\hat \theta \left( t \right)$ obtained with the help of the laws \eqref{eq12}, \eqref{eq27} and errors $\tilde q\left( t \right){\text{, }}\dot {\tilde q}\left( t \right)$ for implementations \eqref{eq23} with \eqref{eq14}, \eqref{eq23} with \eqref{eq27}.

\begin{figure}[h]
      \centering
      \includegraphics[scale=0.6]{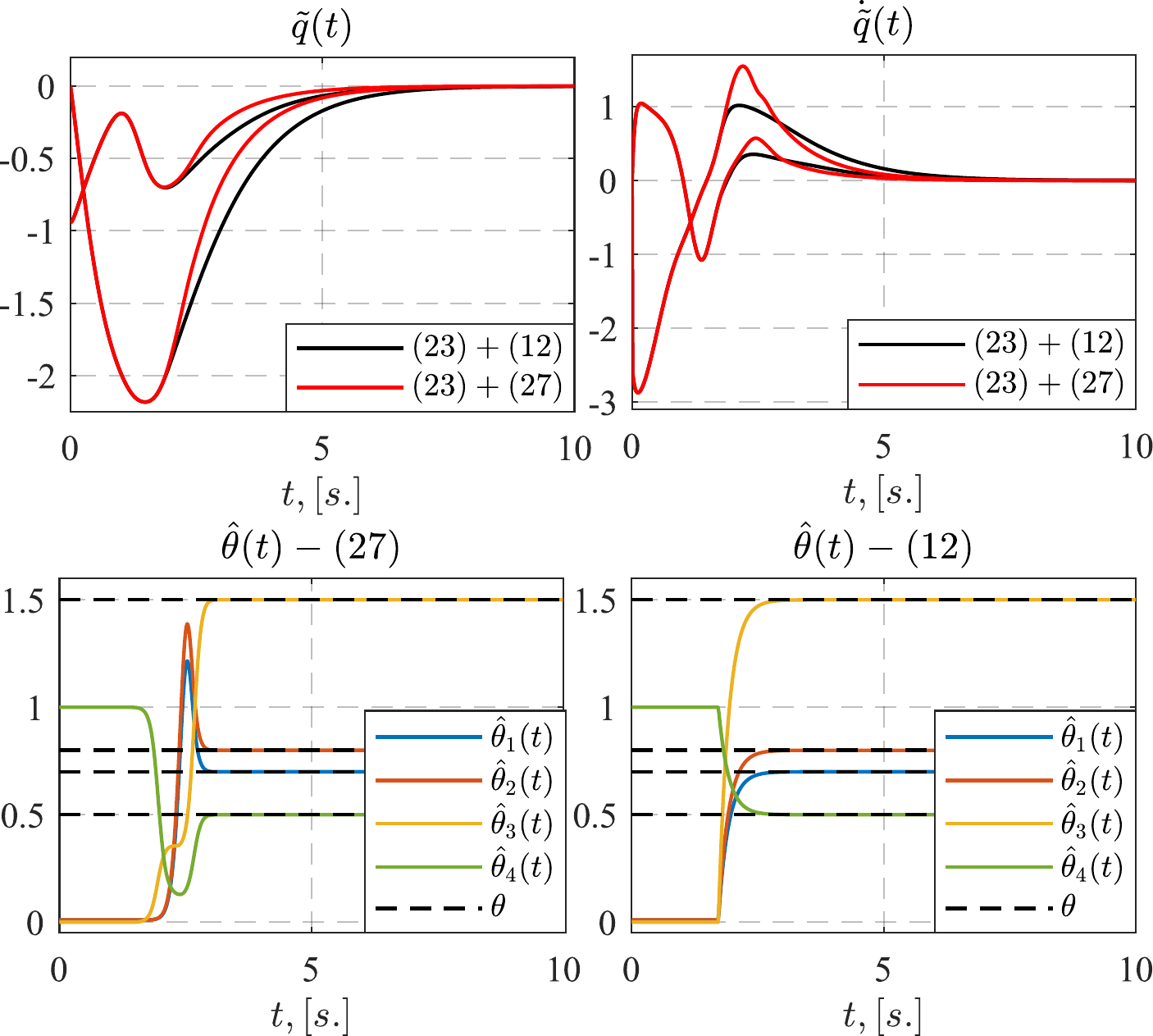}
      \caption{Transient behavior of $\hat \theta \left( t \right)$ and $\tilde q\left( t \right){\text{, }}\dot {\tilde q}\left( t \right)$}
      \label{Figure2} 
      \end{figure}
      
The simulation results confirmed the effectiveness of the proposed estimation law. Owing to the monotonicity of the elements of $\tilde \theta \left( t \right)$, compared to the control law \eqref{eq23} with \eqref{eq27}, the overshoot was reduced for $\dot {\tilde q}\left( t \right)$.

In addition, unlike \eqref{eq27}, the proposed law does not require ({\it i}) special selection of $\hat \theta \left( 0 \right)$ and is implementable under any initial conditions (the law \eqref{eq27} does not allow one to choose ${\eta _1}\left( 0 \right) = 0,{\text{ }}{\eta _2}\left( 0 \right) = 0$ due to the definition of the mapping $\hat \theta \left( 0 \right) = {\mathcal{D}^I}\left( {\hat \eta \left( 0 \right)} \right)$), ({\it ii}) low and upper bounds $\theta_{4}^m,\; \theta_{2}^M,\; \theta_{1}^M,\; \theta_{2}^m$.

\section{Conclusion}
The unknown parameters estimation law for one class of NLPRE was proposed. In contrast to existing solutions, elementwise monotonicity of the parametric error was ensured. Necessary and sufficient implementability conditions for the developed law were: ({\it i}) the regressor finite excitation requirement, ({\it ii}) existence of inverse function from overparameterized parameters to physical ones, ({\it iii}) that only polynomials functions in $\theta$ were used to form overparametrization. The results can be applied to improve the solutions quality of adaptive control and observation problems from recent studies \cite{b9}, \cite{b12}, \cite{b11}, \cite{b5}. 

\bibliographystyle{plain}        % Include this if you use bibtex 
\bibliography{autosam}           % and a bib file to produce the 

\begin{thebibliography}{10}

\bibitem{b4}
S.~Aranovskiy, A.~Bobtsov, R.~Ortega, and A.~Pyrkin.
\newblock Parameters estimation via dynamic regressor extension and mixing.
\newblock In {\em Proc. Amer. Control Conf.}, pages 6971--6976, 2016.

\bibitem{b9}
A.~Bobtsov, R.~Ortega, S.~Aranovskiy, and R.~Cisneros.
\newblock On-line estimation of the parameters of the windmill power
  coefficient.
\newblock {\em Systems \& Control Letters}, 164:105242, 2022.

\bibitem{b12}
R.~Cisneros and R.~Ortega.
\newblock Identification of nonlinearly parameterized nonlinear dissipative
  systems.
\newblock {\em IFAC-PapersOnLine}, 55(12):79--84, 2022.

\bibitem{b15}
A.I. Glushchenko and K.A. Lastochkin.
\newblock Unknown piecewise constant parameters identification with exponential
  rate of convergence.
\newblock {\em Int. J. of Adaptive Control and Signal Proc.}, 37(1):315--346,
  2023.

\bibitem{b14}
A.I. Glushchenko, K.A. Lastochkin, and V.A. Petrov.
\newblock Exponentially stable adaptive control. {P}art {I}. {T}ime-invariant
  plants.
\newblock {\em Autom. and Remote Control}, 83(4):548--578, 2022.

\bibitem{b3}
L.~Ljung.
\newblock {\em System Identiﬁcation: Theory for the User}.
\newblock Prentice Hall, New Jersey, 1987.

\bibitem{b11}
R.~Ortega, A.~Bobtsov, R.~Costa-Castello, and N.~Nikolaev.
\newblock Parameter estimation of two classes of nonlinear systems with
  non-separable nonlinear parameterizations.
\newblock 2022.
\newblock arXiv preprint arXiv:2211.06455,
  \url{https://arxiv.org/abs/2211.06455}.

\bibitem{b6}
R.~Ortega, V.~Gromov, E.~Nuño, A.~Pyrkin, and J.G. Romero.
\newblock Parameter estimation of nonlinearly parameterized regressions:
  application to system identification and adaptive control.
\newblock {\em IFAC-PapersOnLine}, 53(2):1206--1212, 2020.

\bibitem{b5}
R.~Ortega, V.~Gromov, E.~Nuño, A.~Pyrkin, and J.G. Romero.
\newblock Parameter estimation of nonlinearly parameterized regressions without
  overparameterization: application to adaptive control.
\newblock {\em Automatica}, 127:109544, 2021.

\bibitem{b7}
R.~Ortega, J.G. Romero, and S.~Aranovskiy.
\newblock A new least squares parameter estimator for nonlinear regression
  equations with relaxed excitation conditions and forgetting factor.
\newblock {\em Systems \& Control Letters}, 169:105377, 2022.

\bibitem{b8}
A.~Pavlov, A.~Pogromsky, N.~van~de Wouw, and H.~Nijmeijer.
\newblock Convergent dynamics, a tribute to {B}oris {P}avlovich {D}emidovich.
\newblock {\em Systems \& Control Letters}, 52(3):257--261, 2004.

\bibitem{b2}
J.G. Romero, R.~Ortega, and A.~Bobtsov.
\newblock Parameter estimation and adaptive control of euler–lagrange systems
  using the power balance equation parameterisation.
\newblock {\em Int. J. of Control}, pages 1--13, 2021.

\bibitem{b16}
J.E. Slotine and L.~Weiping.
\newblock Adaptive manipulator control: A case study.
\newblock {\em IEEE Trans. on Automatic control}, 33(11):995--1003, 1988.

\end{thebibliography}
                                 % bibliography (preferred). The
                                 % correct style is generated by
                                 % Elsevier at the time of printing.

%\begin{thebibliography}{99}     % Otherwise use the  
                                 % thebibliography environment.
                                 % Insert the full references here.
                                 % See a recent issue of Automatica 
                                 % for the style.
%  \bibitem[Heritage, 1992]{Heritage:92}
%     (1992) {\it The American Heritage. 
%     Dictionary of the American Language.}
%     Houghton Mifflin Company.
%  \bibitem[Able, 1956]{Abl:56}
%     B.~C.~Able (1956). Nucleic acid content of macroscope. 
%     {\it Nature 2}, 7--9. 
%  \bibitem[Able {\em et al.}, 1954]{AbTaRu:54}   
%     B.~C. Able, R.~A. Tagg, and M.~Rush (1954).
%     Enzyme-catalyzed cellular transanimations.
%     In A.~F.~Round, editor, 
%     {\it Advances in Enzymology Vol. 2} (125--247). 
%     New York, Academic Press.
%  \bibitem[R.~Keohane, 1958]{Keo:58}
%     R.~Keohane (1958).
%     {\it Power and Interdependence: 
%     World Politics in Transition.}
%     Boston, Little, Brown \& Co.
%  \bibitem[Powers, 1985]{Pow:85}
%     T.~Powers (1985).
%     Is there a way out?
%     {\it Harpers, June 1985}, 35--47.

%\end{thebibliography}

%\appendix
%\section{A summary of Latin grammar}    % Each appendix must have a short title.
%\section{Some Latin vocabulary}         % Sections and subsections are supported  
                                        % in the appendices.
\end{document}